\documentclass[letterpaper]{article} 
\usepackage{aaai25}  
\usepackage{times}  
\usepackage{helvet}  
\usepackage{courier}  
\usepackage[hyphens]{url}  
\usepackage{graphicx} 
\usepackage{natbib}  
\usepackage{caption} 
\usepackage{algorithm}
\usepackage{algorithmic}

\usepackage{newfloat}
\usepackage{listings}
\DeclareCaptionStyle{ruled}{labelfont=normalfont,labelsep=colon,strut=off} 
\floatstyle{ruled}
\newfloat{listing}{tb}{lst}{}
\floatname{listing}{Listing}
\usepackage[dvipsnames]{xcolor} 
\usepackage{colortbl}
\usepackage{multirow}
\definecolor{myorange2}{rgb}{0.966, 0.881, 0.805}
\definecolor{myblue2}{rgb}{0.854, 0.942, 0.979}

\title{\textit{Hide or Highlight}: Understanding the Impact of Factuality Expression on User Trust}

\author {
    Hyo Jin Do,
    Werner Geyer
}
\affiliations {
    IBM Research, Cambridge, MA, USA\\
    hjdo@ibm.com, 
    werner.geyer@us.ibm.com
}

\usepackage{bibentry}

\begin{document}

\maketitle

\begin{abstract} 
Large language models are known to produce outputs that are plausible but factually incorrect. To prevent people from making erroneous decisions by blindly trusting AI, researchers have explored various ways of communicating factuality estimates in AI-generated outputs to end-users. However, little is known about whether revealing content estimated to be factually incorrect influences users' trust when compared to hiding it altogether. We tested four different ways of disclosing an AI-generated output with factuality assessments: \textit{transparent} (highlights less factual content), \textit{attention} (highlights factual content), \textit{opaque} (removes less factual content), \textit{ambiguity} (makes less factual content vague), and compared them with a \textit{baseline} response without factuality information. We conducted a human subjects research (N=148) using the strategies in question-answering scenarios. We found that the \textit{opaque} and \textit{ambiguity} strategies led to higher trust while maintaining perceived answer quality, compared to the other strategies. We discuss the efficacy of hiding presumably less factual content to build end-user trust.
\end{abstract}

\maketitle

\section{Introduction}~\label{sec:intro}
Large language models (LLMs) are prone to generating factually incorrect information and presenting it as if it were true, a phenomenon known as ``hallucinations''~\cite{ji2023survey}.  
Hallucinations are not only a risk for companies leveraging LLMs in their customer-facing applications, but they can also cause significant harm to end-users. 
For example, a company stock value dropped after its AI-powered product generated a factual error during a public demonstration~\cite{Google_Webb} and another company was sued due to false information given by its AI chatbot~\cite{aircanada_news}. 

Mitigating hallucinations has become an urgent topic in research communities and for AI practitioners. They have developed algorithmic approaches such as 
optimizing LLM outputs by retrieving a knowledge base as a reference~\cite{muther2023citations,lewis2020retrieval,yan2024corrective} or attributing information in the LLM outputs back to the sources~\cite{paes2024multi,lundberg2017unified_SHAP,sundararajan2017axiomatic,ju2023hierarchical}. Along with the algorithmic advances, another critical question is how the outputs of LLMs are \textit{communicated} to end-users in a way that users neither overrely nor underrely on LLM responses. 
Researchers found that calculating and presenting factuality estimates in addition to an AI-generated response, i.e. assessment of how factual the response is, can be an effective way to provide insights into the reliability of the response~\cite{bengio2000neural,farquhar2024detecting}. Factuality estimates allow users to recognize incorrect information in the AI-generated answer, thus helping them to make corrections and better decisions~\cite{cao2024designing}.

A popular approach for presenting factuality estimates is to use visual highlights~\cite{do2024facilitating, vasconcelos2023generation, sun2022investigating,bo2024rely}. For example, \citet{do2024facilitating} have found that incorporating in-line highlighting in LLM responses based on factuality levels made it easier for users to validate the accuracy of the LLM's response and increased their trust, compared to a baseline where no style was applied.
More recently, researchers have begun using the language itself such as epistemic markers to convey factuality~\cite{kim2024m,zhou2024relying,cheong2024not}. For instance, \citet{kim2024m} have shown that the use of hedging language combined with first-person expressions (e.g., ``I’m not sure, but...'') reduced overreliance while improving users' judgments about the correct answer. 
These works assume that the original AI-generated answer is always disclosed to the user, which consists of content estimated to be less factual (i.e. low factuality content) or more factual (i.e. high factuality content). 
In this paper, we challenge this assumption by asking the following question: 
\textit{Can we build user trust by hiding low factuality content from the AI-generated answer?}

The disclosure of factuality estimates along with AI-generated information has been widely acknowledged in traditional AI systems as a means of transparency, allowing users to gain a deeper understanding of the system's capabilities and limitations~\cite{liao2023ai,bhatt2021uncertainty}. However, in human-human conversations, Grice describes how people achieve cooperative communications by obeying the quality maxim in which people do not say things that they believe to be false, and if they do, it can erode trust~\cite{grice1975logic}. In a question-answering scenario, it is often believed that ``incorrect answers are worse than no answers''~\cite {burger2001issues}. 
In the context of human-AI interaction, \citet{vasconcelos2023generation} discovered that participants were confused by AI highlighting potential errors and felt that these should not be visible to them.
Despite these perspectives, there is a lack of empirical research on effects of hiding low factuality content from end-users and its impact on trust. 
In this study, we explore four strategies of communicating an AI-generated response with factuality estimates (low or high) and a baseline: 
\begin{enumerate}
\item \textit{Baseline}: The original AI response is shown without any factuality assessments (e.g., ``He studied architecture before moving to Paris in 1950'')
\item \textit{Transparent}: Low factuality content within the response is highlighted (e.g., ``He studied architecture \colorbox{myorange2}{before} \colorbox{myorange2}{moving to Paris in 1950}'')
\item \textit{Attention}: High factuality content within the response is highlighted (e.g., ``\colorbox{myblue2}{He studied architecture} before moving to Paris in 1950'')
\item \textit{Opaque}: Low factuality content is simply removed (e.g., ``He studied architecture \colorbox{myorange2}{[..]}'')
\item \textit{Ambiguity}: Low factuality content is replaced by vague statements that are not factually incorrect (e.g., ``He studied architecture before moving to another country in the 1950s''). 
\end{enumerate}

In contrast to the \textit{baseline} and the \textit{transparent} designs, the other three designs employ different strategies to downplay or remove low factuality content, specifically through removing the highlights, omitting the content entirely, or replacing the content with ambiguous language. As a result, the \textit{opaque} and \textit{ambiguity} strategies altered some content of the AI-generated response, whereas the other three strategies preserved the original response. 
We aimed to understand the effects of these strategies on user trust in the AI model and perceptions of the AI-generated answer quality and transparency in question-answering scenarios. To investigate these effects, we conducted human subjects research (N=148) in which a participant answered a set of questions after reading an AI-generated answer, where low factuality content was shown or hidden using one of the five strategies.

This research has significant implications for the design of human-AI interactions by exploring the idea of hiding low factuality content in an AI-generated answer to build user trust. The key contributions are as follows:
\begin{enumerate}
    \item We introduce the \textit{ambiguity} strategy, a novel approach that replaces low factuality content with vague statements by reducing the precision of the information. We explain our technical implementation and evaluation of the approach, and outline areas for future improvement. 
    \item 
    Through a user study with 148 participants, each completing 4 tasks (total 592 tasks), we discovered that adopting \textit{opaque} or \textit{ambiguity} strategies that hide low factuality content can enhance user trust while maintaining the perceived answer quality and transparency. 
    \item We discuss the design implications of the proposed strategies, promoting strategies that enhance the accuracy of the AI-generated response over transparency-based approaches.
    Our findings are relevant for AI practitioners and application developers who need to make decisions about the best approach to choose for their use cases. 
\end{enumerate}

\section{Related Work}
Our study is situated at the intersection of three research areas, LLM hallucinations and factuality, communication strategies, and trust in AI systems.

\subsection{LLM Hallucinations and Factuality}
LLMs have shown impressive capabilities in generating human-like natural language responses, but their potential to generate factually incorrect outputs, a phenomenon known as ``hallucinations'', has sparked concerns. 
The counterpart of hallucination is factuality, defined as truthfulness or the quality of being based on facts~\cite{ji2023survey,maynez2020faithfulness}. 
LLMs can present hallucinations with the same confidence as correct information,  which increases the risk of misinforming users who overrely on AI-generated answers~\cite{spatharioti2023comparing}. 
There is also a risk of underreliance on AI, where people might underestimate its potential or be hesitant to accept AI's recommendations, leading to missed opportunities for productivity and performance improvements.
There is an ongoing research effort in the machine learning community to advance LLMs and develop techniques to tackle hallucinations, such as improving the training process, refining the model architecture and parameters, and augmenting LLMs with knowledge bases~\cite{huang2025survey,ji2023survey}. 


Along with the effort to advance the underlying LLM technology, researchers have been seeking ways to enhance transparency of AI models. 
One form of transparency is estimating and expressing factuality cues, which can help users examine how much they should trust information generated by AI~\cite{bhatt2021uncertainty,do2024facilitating,cao2024designing,kim2024m,cheng2024relic}.
However, unique characteristics of LLMs such as generating open-ended, unstructured, and context-dependent texts, introduce new challenges in measuring and communicating factuality.  


For factuality assessment, uncertainty-based metrics are widely used, as factual sentences are likely to contain tokens with lower uncertainty~\cite{manakul2023selfcheckgpt}. 
Other estimation approaches include measuring the consistency between multiple LLM responses~\cite{manakul2023selfcheckgpt,cheng2024relic} or cross-checking claims in the LLM output against retrieved external sources~\cite{min2023factscore,thorne2018fever,wadden2020fact}.
Algorithms for factuality estimation in LLMs is an active topic of research and is beyond the scope of this paper.

After measuring factuality, another critical question is how to communicate factuality that is useful to users~\cite{leiser2023chatgpt}. 
Many researchers have studied the use of highlights to convey factuality and therefore help users analyze the accuracy of the AI-generated output easily and take actions if needed~\cite{do2024facilitating, sun2022investigating, spatharioti2023comparing,bo2024rely}. 
For example, \citet{do2024facilitating} compared various styles of expressing factuality scores, including highlights and numerical scores annotated at the phrase or term level. They found that users trusted and preferred a style that highlights every phrase in the LLM output with varying background colors based on a factuality color scale. 
More recently, researchers have begun using language itself to convey the likelihood of correctness. For instance, \citet{kim2024m} investigated the use of hedging language combined with first-person expressions and found that such expressions reduced over-trust while improving the accuracy of the users' judgments. 
Our work contributes to this research area, aiming to identify the best way to communicate factuality in AI-generated content to users in a way that builds an appropriate level of trust.

\subsection{Communicating Low Factuality Information}~\label{sec:related_work_hiding_strategies}

Prior studies have explored ways to express factuality cues under the assumption that the original AI-generated content is always disclosed to the user even when it is estimated to be less factually correct. As a result, users should calibrate their trust based on the level of factuality presented. In this paper, we challenge this assumption by investigating how to effectively \textit{hide} less factual information after factuality is calculated. 
Grice emphasizes the importance of quality maxim in human-human cooperative conversations which includes, 1) do not say things that you believe to be false and 2) do not say things for which you do not have sufficient evidence~\cite{grice1975logic}. 
In the context of human-AI interaction, \citet{vasconcelos2023generation} highlighted the content that is likely to be edited by users and discovered that users were confused ``why the AI was highlighting its own mistakes'' and ``should be not shown'' instead.

Despite these views, there has been limited empirical research on whether and how to effectively hide low factuality information from end-users and how it influences their trust and perceptions. The closest work we found was done by \citet{wang2021show} in the context of a feature attribution method (e.g., LIME~\cite{ribeiro2016should}), a machine learning technique to explain how influential each measured input feature value is for an output prediction. The authors proposed a method that adjusts the regularization penalty 
on the LIME explainer for suppressing uncertain feature attributions and shifting attributions to other features. The authors found that their suppressing technique can improve user trust in the prediction models. However, it is challenging to apply their suppressing approach to LLMs, which often generate long sequences of text with complex, context-dependent decision-making processes. In this paper, we explored alternative strategies to hide low factuality content in an LLM output and compared them with designs that discloses low factuality content.

\subsection{Trust in AI Systems}
Trust is defined as ``an attitude that an agent will help achieve an individual's goals in a situation characterized by uncertainty and vulnerability''~\cite{lee2004trust}. 
Trust is a crucial factor in determining users' behavioral intentions to adopt technology~\cite{chao2019factors}. 
While trust is an attitude, \textit{reliance} is defined as a user's ``behavior that follows from the advice of the system''~\cite{scharowski2022trust}. 
In automation literature, \textit{compliance} is also used to convey a similar concept, which refers to the ``number of times participants follow the systems' recommendations, both correct and incorrect ones''~\cite{vereschak2021evaluate}.
Calibrating an appropriate level of trust that matches the true capabilities of an AI system is critical to avoid overreliance/overcompliance and underreliance/undercompliance~\cite{lee2004trust}. 

The relationship between trust and transparency is known to be complex. Kizilcec has found that when individuals receive a lower grade than expected, providing some transparency with procedural information fosters trust~\cite{kizilcec2016much}.
\citet{vossing2022designing} have found that trust increases when the AI system provides information on its reasoning, while trust decreases when the AI system provides information on uncertainty. 
In this work, we aim to build trust by increasing the factual accuracy of the AI-generated outputs, even at the cost of reduced transparency.


\section{Strategy Designs}
This paper investigates five strategies for communicating low factually content within an AI-generated answer: \textit{transparent}, \textit{attention}, \textit{opaque}, \textit{ambiguity}, and \textit{baseline}.
In addition to prior research and theories that motivated our designs, we conducted a formative study to inform the designs of the \textit{transparent}, \textit{attention}, and \textit{opaque} strategies, and an experiment to implement the \textit{ambiguity} strategy. 

\subsection{Background and Theoretical Motivation}

Highlights have been widely used and tested as factuality cues~\cite{do2024facilitating, sun2022investigating, spatharioti2023comparing,bo2024rely}.
Prior works have also found that communicating factuality cues is effective in calibrating trust as compared to not communicating them~\cite{do2024facilitating, sun2022investigating, spatharioti2023comparing}. 
Our research builds on these transparency-based approaches by including the \textit{transparent} strategy.

We introduce the \textit{attention} strategy that highlights high factuality content. The strategy was motivated by 
the selective attention theory~\cite{bater2019selective}, which is defined as the cognitive process of attending to fewer sensory stimuli while ignoring or suppressing all other sensory inputs.
While the \textit{attention} strategy doesn't remove or change low factuality content directly, it may subtly influence users' cognitive focus, leading people to steer attention away from low factuality content and focus on high factuality content.

The \textit{opaque} strategy simply removes low factuality content from the answer text. In the case of short answers like a word or a sentence, the \textit{opaque} strategy may simply remove the whole answer and respond as uncertain or unknown. 
A similar study was done by \citet{wester2024ai} that investigated four denial styles of LLMs: 1) factual, which provides a denial followed by a reason, 2) diverting, which the LLM steers away from the request, 3) opinionated, which the response emphasizes the inappropriateness of the request, and 4) baseline, which simply states that the LLM cannot provide assistance. The study found that the diverting denial strategy (e.g., ``While I can't provide specifics on A, I can discuss B'') resulted in lower frustration and higher satisfaction in LLM interactions compared to the baseline denial (e.g., ``I am sorry, but I'm unable to provide the information''). 
However, in the case of long answers, removing the entire answer is not informative to the user. Ideally, incorrect information should be removed from the answer while preserving correct information. This is challenging because removing  parts that are estimated to be less factual may harm the answer quality, such as disrupting the flow of the answer and making it feel incomplete or uninformative, especially when the removed parts are essential.

To address this potential issue of the \textit{opaque} strategy, we propose the \textit{ambiguity} strategy that replaces low factuality content with vague statements.
Through this strategy, we expect that the converted answer is less likely to be incorrect because vagueness reduces the precision level of potentially incorrect details. 
At the same time, this strategy could minimize the impact on the quality of the response by maintaining the structure of the sentence. 
This approach was inspired by  strategic ambiguity~\cite{eisenberg1984ambiguity} in organizational communication, defined as ``intentionally crafting or presenting statements in a way that allows for multiple interpretations or leaves certain details vague or unclear.'' Researchers found that deliberate vagueness in human-human communications helps preserve the speakers' credibility~\cite{williams1975equivocation}.
In machine learning research, Mohri and Hashimoto proposed conformal factuality, a framework that can ensure high-probability correctness guarantees for language models by progressively making the model outputs less specific~\cite{mohri2024language}. 
However, little is known about the effectiveness of the ambiguity strategy in human-LLM interactions. In this work, we explore the utility of this ambiguity as a novel way to present low factuality information.

\subsection{Formative Study for Designs}
The goal for this formative study was to identify an appropriate design of the AI-generated answers for the proposed strategies. 
We interviewed six participants to propose the best design to disclose or hide information with examples for inspiration. 
We found that most participants liked the idea of highlighting rather than alternative options for annotating low factuality content (e.g., italic, strike-through). The initial designs had yellow highlights for both \textit{transparent} and \textit{attention} strategies and participants recommended the use of different colors such as an orange color to indicate low factuality because ``\textit{orange generally indicates that something is wrong} [P4]''. 
Hence, we chose orange and blue highlights from an accessible color palette for the \textit{transparent} and \textit{attention} strategies to indicate low and high factuality. 

We encouraged participants to come up with their own ideas for the design of the \textit{opaque} strategy, along with three example designs we made: removal design that simply removes low factuality content without any indication, full annotation design that indicates removal by `[text removed due to uncertainty]', and simple annotation design that indicates removal by `[..]'. Most participants preferred the simple annotation design because participants generally wanted a minimalist style to avoid distractions but also be aware that something had been removed: ``\textit{In cases where a paragraph is long, the remaining parts after removal might not connect well to each other. Therefore, it makes sense to inform readers that something has been removed}'' [P3].

\subsection{Ambiguity Strategy Implementation}~\label{sec:ambiguity_strategy_implementation}

\begin{figure*}
\includegraphics[width=\textwidth]{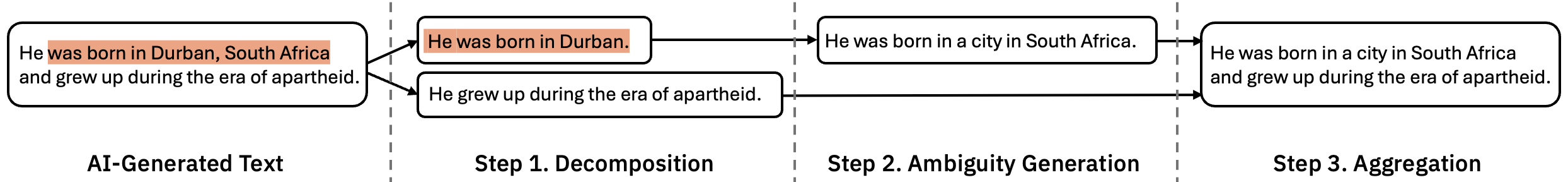}
\caption{Simplified steps for the \textit{ambiguity} strategy implementation. Given the original sentence that contains low factuality content, we conducted the following steps: 1) break down the text into atomic facts, labeling each with factuality estimates (high or low); 2) convert each low factuality fact into an ambiguous fact; and 3) aggregate all facts into a single sentence again.}
\label{fig:ambiguity_steps}
\end{figure*}

The \textit{ambiguity} strategy replaces low factuality content with vague statements that lack precision, making it harder to judge its factual accuracy.
Ideally, the AI-generated answers using the \textit{ambiguity} strategy should not contain incorrect statements, and no highlighting should be needed to indicate low factuality.
To implement the \textit{ambiguity} strategy, we conducted the following steps, as illustrated in Figure~\ref{fig:ambiguity_steps}:
\begin{itemize}
    \item \textit{Step 1. Decomposition}: An LLM breaks down the given text into atomic facts. Each atomic fact is annotated with factuality (low or high). 
    \item \textit{Step 2. Ambiguity Generation}: Each atomic low-factuality fact is converted to an ambiguous fact. 
    \item \textit{Step 3. Aggregation}: All high-factuality facts and ambiguous facts are aggregated into a single sentence.
\end{itemize}

For step 1, we used an existing dataset~\cite{min2023factscore}, which used ChatGPT to generate atomic facts for each sentence in the AI-generated answer, and human annotations to generate factuality labels for each atomic fact. Note that the authors of the dataset also demonstrate an automatic annotation approach using an LLM, but we used human annotations instead for accuracy.
For steps 2 and 3, we used the outputs from an LLM (ibm/granite-13b-instruct-v2\footnote{https://huggingface.co/ibm-granite}) using \textit{zero-shot with context} (e.g., ``..The user will provide a sentence containing a fact. Rephrase the following fact to be vague or ambiguous..") and \textit{aggregation} (e.g., ``..Combine these facts into a single sentence..")  prompts to convert each incorrect fact to an ambiguous statement and aggregate them into a single sentence. To eliminate confounding effects of errors, the final output was manually reviewed and refined by the researchers. 

The choice of the model and the prompt was based on a separate experiment, where we converted the eight biography examples we used in our user study using five prompt types and seven LLMs. 
The LLM outputs were evaluated by an LLM-as-a-judge framework (using meta-llama/llama-3-70b-instruct as an evaluator)~\cite{pan2024human,ashktorab2025evalassist} with the average score of relevance, coherence, conciseness, completeness, and correctness criteria. 
The full details of the experiment including exact prompts we used are described in Appendix~\ref{appendix:ambiguity}.

\section{Method}~\label{sec:method}
The research questions (RQs) addressed in our study are: 
\begin{itemize}
    \item \textit{RQ1. Trust}: How do varying strategies of communicating low factuality content affect user trust in AI? 
    \item \textit{RQ2. Perceived Answer Quality}: How do varying strategies of communicating low factuality content affect the perceived answer quality? 
\end{itemize}

To answer our research questions, we designed a human subject research experiment in which participants reviewed AI-generated responses. We used an online survey to collect the data. 
We focused on question-answering scenarios where factuality is important, making this an appropriate context for our study. In particular, we used a human-labeled biography dataset~\cite{min2023factscore} for designing question and answer pairs. The dataset consists of 183 people entities who have Wikipedia pages. Each data point consists of a question prompt ``Tell me a bio of [entity]'' and an answer generated by ChatGPT. The dataset also includes a list of atomic facts that are included in each AI-generated answer. These atomic facts were generated by an LLM and revised by human annotators. Each atomic fact was then labeled by human annotators as either factually correct or not, which we used for the factuality cues (i.e. high or low factuality). We randomly selected 8 entities from this dataset with diverse nationalities. 
The biography task has been used in similar prior research studying factuality~\cite{min2023factscore, manakul2023selfcheckgpt, cheng2024relic} because the scope is broad, covering diverse nationalities and professions of people. The generated biographies consist of multiple factual statements, rather than debatable or subjective statements, allowing participants to verify factual correctness easily. 

The AI-generated answers consisted of 153.5 words and 7.5 sentences on average. Within an answer, there were 34 atomic facts on average, and 14.75 atomic facts (43\%) were labeled as incorrect. Low factuality parts of the AI-generated answers were visible in the conditions using the \textit{baseline}, \textit{transparent}, and \textit{attention} strategies. These parts were removed or neutralized using \textit{opaque} or \textit{ambiguity} strategies, leaving no low factuality parts in those conditions. Our supplementary material contains the full list of AI-generated answers displayed for each condition used in the study.

Following the guidelines in~\cite{vereschak2021evaluate} and prior works~\cite{kim2024m}, we used the judge-advisor paradigm, which is a common decision-making task to measure trust-related behavior. The paradigm consists of presenting an AI recommendation and giving the participant the option to follow the recommendation~\cite{vereschak2021evaluate}. We used the same setup where the AI model generates an answer to a question and the participants can then decide to accept or reject the AI-generated answer. 
The task showed an AI-generated biography of an entity that contains factual statements. Then, a subsequent question (i.e. ``Using only the information provided in this survey and/or the reference link, please indicate whether you think the following statement is correct or incorrect: [statement]'') asked participants to decide whether to accept (i.e. judge as correct) or reject (i.e. judge as incorrect) a statement relevant to the AI-generated answer. We aimed to control participants' fact-checking ability by restricting them to the provided information. The [statement] was randomly chosen between correct and incorrect labeled atomic facts from the dataset, associated with each AI-generated answer~\cite{min2023factscore}.

\subsection{Participants}
We recruited participants from our company. We aimed to recruit a diverse participant sample in terms of location, job role, and technical expertise. The eligibility criteria required proficiency in English, as all study materials were written in English. 
We advertised and distributed the survey link using an internal messaging platform within our company. Among 156 participants who completed our survey, we filtered out 8 participants who self-rated their English level as `very basic or none' or didn't pass an attention-check question (e.g., Please select the option labeled `Somewhat agree'). As a result, we used data from 148 participants who were assigned to one of the five conditions, each of which employed one of our five strategies: \textit{baseline} (N=27), \textit{transparent} (N=29), \textit{attention} (N=30), \textit{opaque} (N=33), and \textit{ambiguity} (N=29). 

Participants' work locations consisted of 22 unique countries, with the most common being the US (N=82), India (N=18), Canada (N=10), UK (N=7), and Costa Rica (N=5). Job roles spanned a wide array of disciplines, including software development (N=50), engineering (N=26), sales (N=21), design (N=15), customer service (N=13), and finance (N=11). Participants had a range of experience with AI, with some having heard about it from the news, work, friends, or family (N=9), others reporting that they closely follow AI news (N=29), the largest subset reporting some work or educational experience regarding AI (N=93), and others with significant work experience with AI (N=17).

We ensured that participants' prior knowledge about the biography entity did not confound the study outcomes by carefully choosing the biographies that are not categorized as frequent entities from the dataset~\cite{min2023factscore}. As a result, the average rating for entity familiarity was only 1.05 (SD = 0.22) on a 5-point Likert scale. 
A one-way ANOVA revealed that there was no statistically significant difference in the averages across conditions ($p=0.49$), indicating that the potential confounding effect of prior knowledge was removed. 
We followed our company's personal privacy and survey regulations by asking only the approved demographic questions outlined above.

\subsection{Procedure}
After participants signed a consent form, they completed four biography tasks which were randomly selected from our selection of eight biography examples. 
For each task, participants were told to put themselves in the place of a journalist writing a biography of a person, using the information generated by an AI system after a question prompt, ``Tell me a bio of [entity]''. Their task was to read the AI-generated answer, which was presented according to the strategy of the assigned experimental condition. 
The answer also included a reference, which the participant could click to view the Wikipedia page about the person and verify the AI-generated answer if they wanted. They were told that we assume the information in Wikipedia is factually accurate. 

For trust to be an important part of a relationship, individuals must willingly put themselves at risk or 
be susceptible to the actions of others~\cite{lee2004trust}.
Therefore, we introduced vulnerability in the task by explaining the potential negative virtual consequences on their professional reputation when they perform the tasks poorly.
Refer to Figure~\ref{fig:main} in Appendix~\ref{appendix:UI} to read the full text.

After reading each biography, they answered questions regarding their familiarity with the entity, correctness judgment of a factual statement about the entity, and perceived answer quality. 
After completing the four biography tasks, they answered a post-task survey. The survey included the questions about trust, transparency, and collaboration, followed by background questions. 
After completing a 30-min survey, they were compensated with a reward equivalent to 12.5 USD.

\subsection{Key Measures}

\subsubsection{Trust (RQ1)}
We used a multi-item trust questionnaire (5-point Likert scale) in the post-task survey to measure Trust Belief and Trust Intention, adapted from~\cite{mcknight2002impact}. 

\begin{itemize}
    \item \textit{Trust Belief}: Average rating of six items (Cronbach's $\alpha$=0.84) regarding their perceptions about the system's trustworthiness, such as the system’s perceived ability, benevolence, and integrity.
    \item \textit{Trust Intention}: Average rating of four items (Cronbach's $\alpha$=0.80) regarding their desire to use the system.
\end{itemize}

\subsubsection{Trust-related behaviors}
For each biography task, we asked a judgment question: ``Using only the information provided in this survey and/or the reference link, please indicate whether you think the following statement is correct or incorrect: [\textit{statement}].'' (0: Incorrect, 1: Correct). Using participants' responses to this question, we calculated compliance as a trust-related behavioral measure~\cite{vereschak2021evaluate}, which is defined as the probability that participants accept the AI-generated recommendation. In our task, compliance deals with whether participants accept or reject information from the AI-generated biography. 
This allows us to calculate the following, adopted from~\cite{vereschak2021evaluate}:

\begin{itemize}
   \item \textit{Over-compliance}: Incorrect statements are accepted (i.e. judged as correct).
   \item \textit{Under-compliance}: Correct statements are rejected (i.e. judged as incorrect). 
  \item \textit{Appropriate Compliance}: Correct statements are accepted and incorrect statements are rejected. The calibration of trust can be measured by observing how an individual's behavior aligns with the ideal behavior~\cite{wischnewski2023measuring}, which is the appropriate compliance in our case. 
\end{itemize}


 If a user decides to follow the AI-generated output without cross-checking with the reference, this may suggest that the user trusts AI's suggestions~\cite{kim2024m}. Therefore, we calculated the following:
\begin{itemize}
    \item \textit{Use Link}: Participants' clicks of the Wikipedia reference link we provided (0: not clicked, 1: clicked). 
\end{itemize}

\subsubsection{Trust-related perceptions}
We measured perceived transparency, adopted from~\cite{kim2024m}, which can positively influence user trust. 

\begin{itemize}
    \item \textit{Transparency}: Two individual ratings on each question, 1) ``I feel I had a good understanding of what the AI model A's answers were based on.''; 2) ``I feel I had a good understanding of when the AI model A's answers might be wrong.'' (1: Strongly disagree -- 5: Strongly agree). We found that the ratings of the two questions were not consistent (Cronbach's $\alpha$=0.42), therefore, we used individual ratings for the analysis, rather than averaging them.
\end{itemize}

Participants use AI-generated answers to make judgments rather than deciding on their own, which can be considered an act of human-AI collaboration~\cite{do2024facilitating}. We measured this collaboration experience as research has shown trust is correlated with collaboration experience~\cite{tschannen2001collaboration,vossing2022designing}.

\begin{itemize}
    \item \textit{Human-AI Collaboration}: Rating on the question, ``How would you rate your overall experience of collaborating with the AI model A in performing the tasks?'' (1: Extremely negative -- 5: Extremely positive)
\end{itemize}

\subsubsection{Perceived answer quality (RQ2)}
For each biography task, we asked participants to rate the overall quality of the AI-generated answer on a scale from 1 (worst) to 5 (best) using the following criteria, adopted from prior works~\cite{hirschman2001natural,fabbri2021summeval}: 
    \begin{itemize}
        \item \textit{Correctness}: The answer is factually correct
        \item \textit{Relevance}: The answer is a response to the question
        \item \textit{Conciseness}: The answer does not contain extraneous information
        \item \textit{Completeness}: The answer is complete, not partial
        \item \textit{Coherence}: All the sentences in the answer collectively fit together and sound natural
    \end{itemize}



\subsubsection{Covariates}
We measured AI familiarity, incorrect proportion of the answer, and entity familiarity, to control for potential confounding effects. 

\begin{itemize}
    \item \textit{AI Familiarity}: 
    Rating on the question,
    ``How confident are you in your ability to use artificial intelligence (AI) systems effectively for your tasks, whether for work or personal use? (1: Extremely unconfident -- 5: Extremely confident)
    \item \textit{Incorrect Proportion}: 
    The proportion of factual incorrectness within the AI answer using ``the number of atomic facts labeled as incorrect'' divided by ``the total number of atomic facts within an AI-generated answer''. 
    Specifically, we calculated the proportion before any manipulation (e.g., \textit{opaque} and \textit{ambiguity} strategies). This is because the incorrect proportion may still be noticeable, either through the number of [..] indicators, vague statements, or by the length or incompleteness of the response itself, potentially influencing the dependent variables.
    \item \textit{Entity Familiarity}: Rating on the question, ``How familiar were you with [entity] before reading this text?'' (1: Not familiar at all -- 5: Extremely familiar)
\end{itemize}

\subsection{Statistical Analysis}
We conducted statistical analyses to test the main effect of the experimental conditions  (\textit{baseline}, \textit{transparent}, \textit{attention}, \textit{opaque}, \textit{ambiguity}). We added AI Familiarity and Incorrect Proportion as covariates in all analyses. 
We reported p-values less than 0.05 to determine statistical significance. 

For dependent variables (DVs) that were measured once in the post-task survey (Trust Belief, Trust Intention, Transparency), we performed an analysis of variance (ANOVA) to gain p-values (e.g., DV $\sim$ Condition + AI Familiarity + Incorrect Proportion). If the effect was significant, we conducted post hoc Tukey HSD tests for pairwise comparisons. 
For Human-AI Collaboration measure, the Fligner-Killeen test showed a violation of homogeneity of variances ($p<.05$). Therefore, we conducted a non-parametric Kruskal-Wallis test instead, followed by post hoc Wilcoxon Rank Sum tests with Bonferroni corrections.

For dependent variables that were repeatedly measured for every biography task (Compliance,  Use Link, Perceived Answer Quality), we built linear mixed models (LMMs). 
We used the experimental condition as a fixed-effect factor, and the task ID and the participant ID as random-effect factors (e.g., DV $\sim$ Condition + AI Familiarity + Incorrect Proportion + (1$\mid$Task) + (1$\mid$Participant)). If significant, we conducted post hoc tests for pairwise comparisons with Bonferroni corrections. We obtained p-values using likelihood-ratio chi-squared tests of the full model with the effect in question against the model without the effect in question, which is a common method for LMMs~\cite{winter2013very}.


\section{Results}
\setlength{\tabcolsep}{1mm}  
\begin{table*}[]
\centering
\scriptsize	
\begin{tabular}{c|c|c|c|c|c|c|c}
\hline
RQs & DVs & Baseline & Transparent & Attention & Opaque & Ambiguity  & Significant Differences ($p<.05$) \\ 
\hline
\multirow{8}{*}{Trust} & Trust Belief & 3.22 (1.02) & 3.23 (0.95)& 3.31 (0.84)& 3.87 (0.45)&3.87 (0.68) & 
\begin{tabular}{@{}c@{}}
\textit{Transparent, Attention, Baseline} $<$ \textit{Opaque, Ambiguity}
\end{tabular}\\\cline{2-8}
 & Trust Intention & 2.26 (1.02)& 2.46 (1.09)& 2.48 (1.07)& 2.83 (0.66)&2.90 (0.98)  & -\\\cline{2-8}
& Appropriate Compliance (\%)& 0.78 (0.25) & 0.66 (0.32)& 0.78 (0.26)& 0.86 (0.2)& 0.9 (0.16)  &
\begin{tabular}{@{}c@{}}
\textit{Transparent} $<$ \textit{Opaque, Ambiguity} 
\end{tabular}\\\cline{2-8}
 & Use Link (\%) & 0.58 (0.42)& 0.37 (0.44)& 0.60 (0.43)& 0.59 (0.45)& 0.56 (0.46)  & - \\
 \cline{2-8}
   & Transparency Q1 & 4.04 (0.98)& 3.59 (0.82)& 3.57 (0.97)& 3.85 (0.83) & 4.03 (0.73)  & - \\\cline{2-8}
 & Transparency Q2 & 3.26 (1.13)& 3.69 (0.89)& 3.37 (1.19)& 3.12 (1.11)& 2.97 (1.24) & - \\\cline{2-8}
& Human-AI Collaboration  &  3.15 (1.17)& 3.41 (1.02)& 3.13 (1.04) & 3.82 (0.64) & 3.79 (0.73) & \textit{Attention} $<$\textit{Opaque}\\
\hline
\multirow{5}{*}{\begin{tabular}{@{}c@{}}Perceived\\Answer\\Quality\end{tabular}} & Correctness & 3.15 (1.09)& 2.78 (0.76)& 3.17 (0.85)& 3.77 (0.57)& 3.78 (0.71)  & 
\begin{tabular}{@{}c@{}}
\textit{Transparent, Attention, Baseline} $<$  \textit{Opaque, Ambiguity}
\end{tabular}\\\cline{2-8}
& Relevance & 4.15 (0.80)& 3.84 (0.80)& 4.15 (0.71)& 4.07 (0.69)& 4.27 (0.68)  &  - \\\cline{2-8} 
& Conciseness & 3.91 (0.82)& 3.54 (0.75)& 3.87 (0.69)& 3.98 (0.65)& 3.88 (0.81)  & - \\\cline{2-8} 
& Completeness & 3.45 (0.99)& 3.39 (0.89)& 3.61 (0.86)& 3.07 (0.87)& 3.43 (0.91)  & - \\\cline{2-8} 
& Coherence & 3.97 (0.83)& 3.77 (0.76)& 4.20 (0.64)& 3.70 (0.83)& 4.00 (0.86)  & - \\
\hline
\end{tabular}
\caption{Descriptive statistics of key dependent variables (DVs) for all conditions. We reported the means and standard deviations in parentheses. All DVs were measured with a 5-point Likert scale, except Appropriate Compliance and Use Link which used the decimal equivalent of proportions (\%).}~\label{tab:descriptive}
\end{table*}

We report how the strategies affect end-user trust (RQ1). Additionally, we describe how these strategies impact the perceived quality of AI-generated answers (RQ2). Descriptive statistics for all key dependent variables are summarized in Table~\ref{tab:descriptive}. 

\subsection{RQ1. Trust}

\subsubsection{Trust Belief and Trust Intention}
There was a significant effect of condition on the means of Trust Belief ($F(4,141)=5.89, p<.01$, $\eta_{p}^{2} = 0.14$ with 95\% $CI = [0.04, 0.24]$). 
Post hoc pairwise comparisons showed that participants in the \textit{opaque} condition had significantly higher Trust Belief ratings compared to those in the \textit{attention} condition ($p<.05$), the \textit{transparent} condition ($p<.05$), and the baseline condition ($p<.05$). Similarly, participants in the \textit{ambiguity} condition had significantly higher Trust Belief ratings compared to those in the \textit{attention} condition ($p<.05$), the \textit{transparent} condition ($p<.05$), and the baseline condition ($p<.05$). 

We didn't find a significant effect of condition on the means of Trust Intention ($F(4,141)=2.11$, $p=.05$, $\eta_{p}^{2} = 0.06$ with 95\% $CI = [0, 0.13]$). A possible explanation is that participants were exposed to the intervention for only a short time, which may have changed their perceptions of the AI model, but not have been sufficient to generate a desire to use the system.

\subsubsection{Compliance}
There was a significant effect of condition on Appropriate Compliance (${\chi}^2(4)=18.08, p<.01$, $\eta_{p}^{2} = 0.12$ with 95\% $CI = [0.02, 0.20]$).
Post hoc pairwise comparisons showed that participants in the \textit{opaque} condition had significantly higher Appropriate Compliance compared to those in the \textit{transparent} condition ($p<.01$). We also found that participants in the \textit{ambiguity} condition had significantly higher Appropriate Compliance compared to those in the \textit{transparent} condition ($p<.01$). These results imply that the \textit{opaque} and \textit{ambiguity} strategies are more capable of calibrating trust than the \textit{transparent} strategy.

\begin{figure*}[t]
    \begin{minipage}[t]{.49\textwidth}
        \centering
        \includegraphics[width=0.9\textwidth]{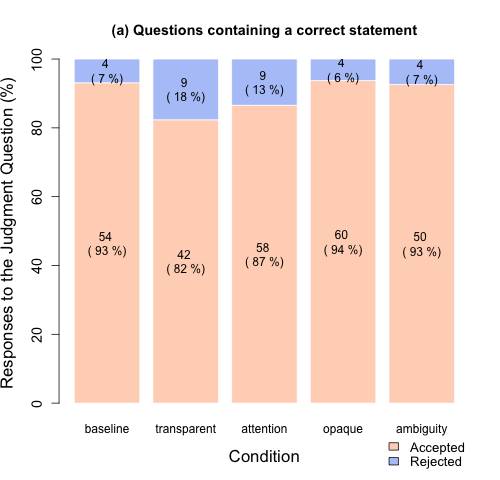}
    \end{minipage}
    \hfill
    \begin{minipage}[t]{.49\textwidth}
        \centering
        \includegraphics[width=0.9\textwidth]{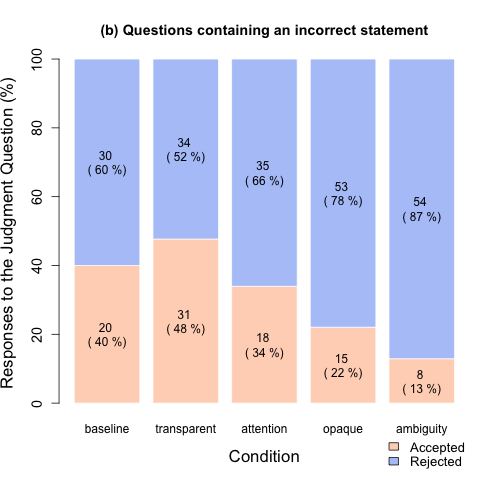}
    \end{minipage}  
    \caption{After reading an AI-generated biography, participants were presented with a statement, either (a) correct or (b) incorrect, and asked whether they accept or reject it. The numbers and proportions of their judgments are shown above in 100\% stacked bar charts, which summarize Appropriate Compliance (correct statement/accepted and incorrect statement/rejected), Under-compliance (correct statement/rejected), and Over-compliance (incorrect statement/accepted) cases.}~\label{fig:compliance}
\end{figure*}

We performed additional analysis by calculating over-compliance with responses to the judgment question containing incorrect statements, and under-compliance with responses to the judgment question containing correct statements. 
In the case of over-compliance, results showed that there was a significant effect of condition on over-compliance (${\chi}^2(4)=13.29, p<.01$
, $\eta_{p}^{2} = 0.09$ with 95\% $CI = [0.01, 0.18]$).
Pairwise comparisons showed that participants in the \textit{ambiguity} condition had significantly lower over-compliance compared to those in the \textit{transparent} condition ($p<.01$). 
We didn't find a significant effect of condition on under-compliance (${\chi}^2(4)=6.63, p=.16$, $\eta_{p}^{2} = 0.06$ with 95\% $CI = [0, 0.13]$).
Figure~\ref{fig:compliance} visualizes the numbers of correct and incorrect statements as well as the proportions of participants' judgments across conditions.


\subsubsection{Use Link}

We didn't find a significant effect of condition on the usage of reference links  (${\chi}^2(4)=5.97$, $p=0.20$, $\eta_{p}^{2} = 0.04$ with 95\% $CI = [0, 0.10]$). 
This aligns with prior findings~\cite{kim2024m} that there is a lack of evidence that the AI factuality expressions affect people's source usage. 

\subsubsection{Transparency}
For Q1 (``I feel I had a good understanding of what the AI model A's answers were based on.''), we didn't find a significant effect of condition on the ratings ($F(4,141)=2.15$, $p = 0.08$, $\eta_{p}^{2} = 0.06$ with 95\% $CI = [0, 0.13]$).
For Q2 (``I feel I had a good understanding of when the AI model A's answers might be wrong.''), we also didn't find a significant effect of condition on the ratings ($F(4,141)=1.74$, $p = 0.14$, $\eta_{p}^{2} = 0.05$ with 95\% $CI = [0, 0.11]$). 
Although we anticipated that communicating factuality would enhance transparency~\cite{liao2023ai,bhatt2021uncertainty}, our study showed no significant effect on transparency ratings. If the goal is to improve user understanding of the AI's decision-making processes and limitations through transparency, our findings suggest that highlighting alone may not be sufficient.

\subsubsection{Human-AI Collaboration}
There was a significant effect of condition on the ratings of the human-AI collaboration (${\chi}^2(4)=13.48, p<.01$, $\eta_{H}^{2} = 0.07$ with 95\% $CI = [0.01, 0.21]$). 
Pairwise comparisons showed that participants in the \textit{opaque} condition rated the human-AI collaboration experience significantly more positive compared to those in the \textit{attention} condition ($p<.05$). This may imply that the AI answers shown in the \textit{opaque} condition were perceived to be more helpful in making correctness judgments about factual statements than the answers in the \textit{attention} condition.

\subsection{RQ2. Perceived Answer Quality}

\subsubsection{Correctness}
There was a significant effect of condition on the ratings of the perceived correctness in AI-generated answers (${\chi}^2(4)=33.69, p<.01$, $\eta_{p}^{2} = 0.20$ with 95\% $CI = [0.09, 0.30]$).
Pairwise comparisons showed that participants in the \textit{opaque} condition rated Correctness significantly higher than those in the \textit{baseline} condition ($p<.05$), the \textit{transparent} condition ($p<.01$) and the \textit{attention} condition ($p<.05$).
Similarly, participants in the \textit{ambiguity} condition rated Correctness significantly higher than to those in the \textit{baseline} condition ($p<.05$), the \textit{transparent} condition ($p<.01$) and the \textit{attention} condition ($p<.05$).
The \textit{opaque} strategy hides incorrect statements and the \textit{ambiguity} strategy converts incorrect statements to neutral ones, whereas the other three conditions disclose incorrect statements. 
Participants were able to accurately assess the correctness of the AI-generated answer, which aligned with our strategy goals.

\subsubsection{Relevance, Conciseness, Completeness, and Coherence}
We didn't find significant effects of condition on the ratings of other quality measures, including Relevance (${\chi}^2(4)=4.87$, $p = 0.30$, $\eta_{p}^{2} = 0.03$ with 95\% $CI = [0, 0.08]$), Conciseness (${\chi}^2(4)=6.33$, $p = 0.18$, $\eta_{p}^{2} = 0.04$ with 95\% $CI = [0, 0.10]$), Completeness (${\chi}^2(4)=6.04$, $p = 0.20$, $\eta_{p}^{2} = 0.04$ with 95\% $CI = [0, 0.10]$), and Coherence (${\chi}^2(4)=7.63$, $p = 0.11$, $\eta_{p}^{2} = 0.05$ with 95\% $CI = [0, 0.11]$). This implies that participants perceived the answer quality as similar across strategies.
Given that participants were aware that some content had been removed from the original answer in the \textit{opaque} condition, we anticipated that the \textit{opaque} strategy would be perceived as less complete or coherent. However, this was not the case. Our data suggests that both the \textit{opaque} and \textit{ambiguity} strategies can maintain the same level of perceived quality as \textit{baseline}, \textit{transparent} and \textit{attention} strategies.


\section{Discussion}

\subsection{Design Implications}
Previous research has often focused on conveying factuality cues by intentionally making the AI appear less confident about its outcomes when it hallucinates. While this approach aims to provide a more accurate and transparent representation of the AI's capabilities, it can also erode users' trust and potentially lead to abandonment of the system. In contrast, our study introduced an alternative design strategy: \textbf{hiding information estimated to be less factual, rather than highlighting it}. 
Our study results show that both the \textit{opaque} and \textit{ambiguity} strategies for managing low factuality content can be promising approaches for AI practitioners and application designers to improve end-user trust, such as improving the perceptions about the AI model's trustworthiness and leading to appropriate compliance with the AI. 

Moreover, participants rated the perceived correctness of the AI-generated answer significantly higher in the \textit{opaque} and \textit{ambiguity} conditions, compared to the \textit{baseline}, \textit{transparent}, and \textit{attention} conditions. Hiding content estimated to be less factual, either by removing it or making it vague, increased the actual correctness of the answer and consequently led to higher perceived accuracy. In reality, factuality estimates may be inaccurate, and future research should explore whether people can accurately evaluate the correctness of the answer with incorrect estimation of factuality. 

We found that the other perceived qualities of the AI answer (relevance, completeness, coherence, conciseness) are rated similarly across conditions, contrary to our expectations that the \textit{opaque} and \textit{ambiguity} strategies may harm the quality of the overall answer. One possible explanation is that people were not familiar with the biography entities, so removing some information from the text did not affect their perceptions significantly. If they had known that some critical information was missing, they might have perceived the answer quality as lower. While our findings imply that designs that remove or obscure low factuality content are preferable, we encourage future research to explore various scenarios involving different levels of familiarity with the topic, as well as various types of LLM responses.

Prior research on AI communicated factuality estimates as a way to achieve transparency~\cite{liao2023ai,bhatt2021uncertainty}. However, our findings suggest that the effect may not apply to LLMs, as the effects of the transparent design were comparable to those of the baseline. One possible reason is that LLMs have complex new capabilities, and simply annotating LLM outputs with factuality cues may not be enough to improve users' understanding of the LLM's decision-making processes and limitations. 

Moreover, we did not find significant differences in the number of times participants clicked on a reliable source (e.g., Wikipedia) to verify AI-generated answers. Researchers have attempted to inform users about the potential errors inherent in LLMs, hoping that this would encourage them to verify information produced by LLMs with reliable sources (e.g., \cite{do2024facilitating,vasconcelos2023generation,bhatt2021uncertainty,cao2024designing,kim2024m}). Similarly, some LLM interfaces incorporate disclaimers about the possibility of hallucinations and encourage validation, such as ChatGPT stating "ChatGPT can make mistakes. Check important info." However, our results suggest that simply informing users about factuality estimates may not be enough to improve their understanding of AI model capabilities or guide users to take action. More advanced designs for LLM transparency, such as interactive visualizations that attribute low factuality content to grounding sources~\cite{do2025understanding,cheng2024relic}, is necessary to help users' understanding and encourage critical thinking.

We also found that the \textit{attention} strategy was not an effective design to hide low factuality information, leading to a negative human-AI collaboration experience. One possible explanation for this result is that overly highlighting high factuality parts of the answer can come across as an attempt to manipulate users' perceptions, which can lead people to view the \textit{attention} strategy more negatively. Additionally, the high factuality highlights might have set high user expectations, which can lead to negative perceptions of the AI's trustworthiness when the AI-generated outcome doesn't meet the inflated expectations~\cite{burgoon2016application}.


\subsection{Practical Implication of the Ambiguity Strategy}
The \textit{ambiguity} strategy accommodates varying interpretations without committing to precise but incorrect assertions. 
It diverges from many real-world guidelines and natural language processing research that focus on generating clear, non-ambiguous texts. For example, Wikipedia offers a manual for editors to look out for and avoid vague or ambiguous claims, which are tagged as `weasel words'\footnote{https://w.wiki/7XCA}. Using a Wikipedia corpus with weasel word annotations, \citet{ganter2009finding} proposed an automatic detection of sentences containing linguistic hedges. 
While clear and precise statements should be prioritized over ambiguous ones to avoid misinterpretation, our experiment suggests that the \textit{ambiguity} strategy is preferable to presenting low factuality information when accurate details are not readily available. 

Implementing the \textit{ambiguity} strategy involves a series of steps, each of which is the focus of active and ongoing research: decomposition into atomic facts~\cite{yan2024atomic, min2023factscore, wright2022generating, chen2022generating}, annotation of factuality~\cite{min2023factscore,kadavath2022language,manakul2023selfcheckgpt}, ambiguity generation~\cite{mohri2024language,angelopoulos2022conformal}, and aggregation~\cite{geva2019discofuse,barzilay2005sentence,narayan2018don}. In this study, we implemented these steps using prior research~\cite{min2023factscore} and LLMs. We compared multiple LLMs and prompts to implement our strategy effectively as explained in Appendix~\ref{appendix:ambiguity}.
While our approach shows promise, there could be alternative ways to implement the strategy better. We call for more research on how to best implement these strategies. 

Moreover, selecting the appropriate level of abstraction presents a challenge in implementing the ambiguity generation step. The granularity of abstraction must be carefully calibrated according to the degree of factual accuracy, as well as the context in which the information will be used. For instance, when some content is completely inaccurate compared to somewhat inaccurate, a higher level of abstraction may be needed to effectively remove incorrect details.
The level of abstraction also depends on the stakeholders' requirements for decision-making, which highlights the need for a more granular, domain-specific approach when applying the ambiguity method. 
For example, in medical research, a high level of precision may be necessary to make health-related decisions, therefore the abstraction should be removed or minimal. 
We encourage future research to refine and expand upon our ambiguity approach and deepen the insights of our exploratory study.

\subsection{Generalizability}
Building trust in an AI system is generally desired for its widespread adoption and effective usage~\cite{chao2019factors}. Without trust, people are less likely to use the technology, which limits the potential benefits of AI such as productivity gains. 
In our study, we focused on a work scenario where participants took on the role of journalists tasked with writing a biography. The AI was designed to estimate factuality on its outputs, assisting users in identifying accurate information by either annotating or hiding less factual parts. Thus, building trust is desired in our scenario, as the AI system can help participants complete their tasks efficiently and accurately with less effort in validating the facts.
However, building trust in AI systems is not always desirable and requires caution. Our findings do not apply to situations involving flawed or biased AI systems, or high-stakes decision-making systems where human oversight and validation are essential.

While our study focused on a question-answering task, the findings and implications may generalize to other information-seeking tasks where accurate information is valued, such as summarization, analysis, and learning tasks. 
Communication of factuality should be tailored to the specific needs and context of different users and applications.
For instance, in a programming scenario, a developer might want to assess the code that is likely to be inaccurate to identify and correct potential errors in their code. In low-stakes contexts, such as movie recommendation systems, recommendations with low confidence may not pose risks to users and can even benefit them by providing access to a broader range of options. 
A hybrid approach that incorporates interactive design elements, such as allowing participants to access hidden information when hovering their mouse over the AI-generated answer, presents a promising pathway forward. 
This approach could serve as a compromise for general-purpose LLMs that are not limited to specific tasks, contexts, or user needs.

\subsection{Limitations and Future Work}
While we assumed in our experiment that the factuality estimation is accurate, 
factuality estimation remains an evolving research area, and it is possible that our strategies may hide parts that are actually correct in real-world situations. This could result in the omission of potentially valuable information. In future research, exploring a more granular scale of factuality estimates could help adjust the amount of details to be either removed or converted. 
Our work focused on a biography task, a widely used and verified task in previous research examining factuality in LLMs. More research should be done on how to estimate and express factuality in other tasks and contexts. 
While we made efforts to recruit participants with diverse backgrounds, including job roles, locations, and technical expertise, our recruitment was restricted to people within our company. Future studies should expand to include more participants with different backgrounds  such as people who have limited or no prior exposure to LLMs.

\section{Conclusion}
Managing hallucinations and establishing trust with generative AI systems are key challenges for the successful adoption of AI applications. Our research in this paper contributes to the large body of research on factuality in AI systems and introduces a novel,  contrarian approach to communicating LLM outputs with low factuality scores. 
We explored five strategies for factuality expression, including the three ``hiding'' strategies, and
our experiment demonstrated that both \textit{opaque} and \textit{ambiguity} strategies were more effective in calibrating trust than the baseline or strategies highlighting either high or low factuality content. 
We found that removing low factuality content or making it ambiguous did not negatively impact the overall perceived quality of AI-generated responses and positively influenced participants' perception of correctness, making this a promising strategy in practice. Hence, we believe that our research has important implications for the design of factuality communication in AI systems. 
For example, future interaction designs can develop novel interaction approaches for dynamic disclosure of information that is likely to be less factual, without giving up on potentially valuable or critical information. 


\bibliography{reference.bib}

\appendix

\onecolumn
\renewcommand\thefigure{\thesection.\arabic{figure}}
\renewcommand\thetable{\thesection.\arabic{table}}
\setcounter{figure}{0}
\setcounter{table}{0}
\section{User Study Interface}~\label{appendix:UI}

\begin{figure*}[ht!]
\centering
\includegraphics[width=0.7\linewidth]{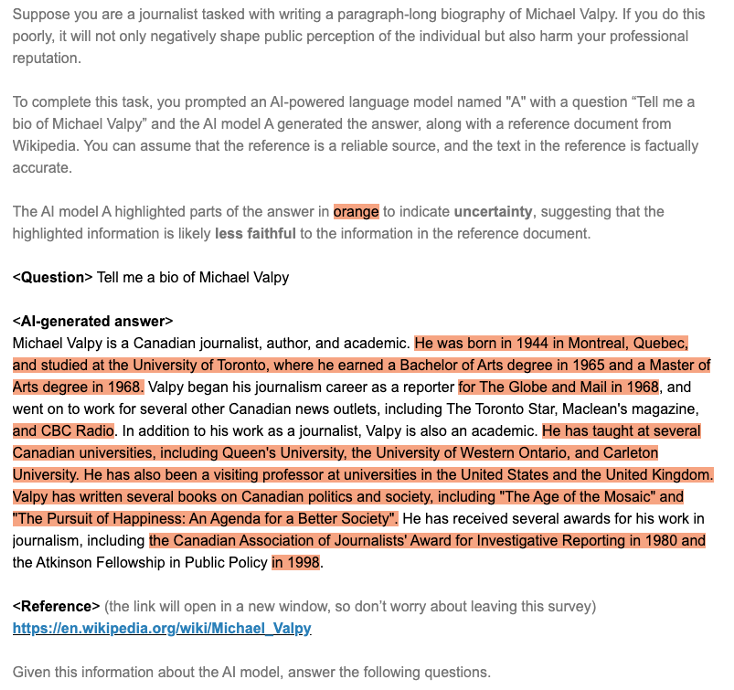}
\caption{A screenshot of the instruction and biography text under the \textit{transparent} condition.} 
\label{fig:main}
\end{figure*}

\newpage

\section{Ambiguity Strategy Implementation and Experiment}~\label{appendix:ambiguity}
Using prompt-engineering guidelines for LLMs\footnote{ OpenAI: https://platform.openai.com/docs/guides/prompt-engineering \\Llama: https://llama.meta.com/docs/model-cards-and-prompt-formats/meta-llama-3/ \\Granite: https://dataplatform.cloud.ibm.com/docs/content/wsj/analyze-data/fm-models-ibm-chat.html?context=wx}, we designed and tested five prompts to implement the ambiguity approach: 
\begin{itemize}
    \item \textbf{P1}: Zero-shot prompt (step 2) + Aggregation (step 3)
    \item \textbf{P2}: Zero-shot prompt with context (step 2) + Aggregation (Step 3)
    \item \textbf{P3}: Few-shot prompt (step 2) + Aggregation (step 3)
    \item \textbf{P4}: Few-shot prompt with context (step 2) + Aggregation (step 3)
    \item \textbf{P5}: Iterative prompt that combines step 2 and step 3
\end{itemize}
See Table~\ref{tab:4prompts_step2} and Table~\ref{tab:aggregation} for the first four prompts, and Table~\ref{tab:prompt_combined} for the iterative prompt.
For each prompt type, we tested seven LLMs, gpt-4o-mini, ibm/granite-13b-instruct-v2, meta-llama/llama-3-8b-instruct, meta-llama/llama-3-70b-instruct, google/flan-t5-xxl, google/flan-ul2, mistralai/mixtral-8x7b-instruct-v01.

Using the eight biography question-answering text we used for the user study, we tested one of the prompt type and an LLM to generate an output with ambiguity. The outputs were then evaluated by an LLM-as-a-judge (i.e. meta-llama/llama-3-70b-instruct)~\cite{pan2024human} with the criteria of relevance, coherence, conciseness, completeness, and correctness, which are listed in Table~\ref{tab:evaluation_prompt}. As a result, we found that the Zero-shot prompt with context (step 2), followed by the Aggregation prompt (step 3), using IBM granite model showed the highest average score of the evaluation criteria, as shown in the Table~\ref{tab:prompt_evaluation}. 
For the user study, we used this best performing prompt type and LLM to generate ambiguity designs. 

Due to the imperfectness of the technology, we further manually edited the generated outputs that were not converted properly. The manual edits followed the ambiguity patterns below, guided by the prior work~\cite{ganter2009finding} and Wikipedia manuals about ambiguity\footnote{https://w.wiki/7XCA}:
\begin{itemize}
    \item Broaden the scope: e.g. in New York → in the US
    \item Underspecify: e.g. numerically underspecified subjects e.g. some people, experts, many
    \item Euphemism: e.g. retire → step away
    \item Approximate: e.g. in 2004 → around 2004, in the 2000s
    \item Passive constructions: e.g. it is believed, it is considered
    \item Use of adverbs: e.g. often, probably
\end{itemize}

\begin{table*}[h]
\resizebox{\textwidth}{!}{ 
\small
\begin{tabular}{c|c|c|c|c|c|c|c}
\hline
Prompt & gpt-4o-mini & granite-13b-instruct-v2 & llama-3-8b-instruct & llama-3-70b-instruct & flan-t5-xxl & flan-ul2 & mixtral-8x7b-instruct-v01 \\ \hline
P1 & 4.075 & 3.375 & 4.025 & 3.775 & 3.2 & 3.3 & 4.2 \\ \hline
P2 & 4.025 & \cellcolor{green!25}{4.375} & 4.025 & 4.025 & 4 & 3.6 & 4.275 \\ \hline
P3 & 4.075 & 4.175 & 4.025 & 3.925 & 3.925 & 3.475 & 3.3 \\ \hline
P4 & 3.775 & 4.2 & 4.225 & 3.85 & 4 & 3.925 & 3.475 \\ \hline
P5 & 3.9 & 2.6 & 3.125 & 3.875 & 2.775 & 2.875 & 3.575 \\ \hline
\end{tabular}
}
\caption{Evaluation results for each prompt type and LLM. The highest score is highlighted in green.}~\label{tab:prompt_evaluation}
\end{table*}

\begin{table*}[h]
\small
\resizebox{\textwidth}{!}{ 
\begin{tabular}{p{\textwidth}}
\hline
\textbf{Step 2. Zero-shot prompt} \\\hline
You are a highly intelligent assistant. You will carefully follow the instructions. Rephrase the following fact to be ambiguous, allowing for various interpretations or obscuring specific details. Your response should only include the answer. Do not provide any further explanation. 
\\ Fact: \{\{incorrect-fact\}\}
\\ Answer: 
\\\hline
\textbf{Step 2. Zero-shot prompt with context} \\\hline
You are a highly intelligent assistant. You will carefully follow the instructions. \\The user will provide a sentence containing a fact. Rephrase the following fact to be vague or ambiguous, allowing for various interpretations or obscuring specific details. Your answer should be still relevant to the sentence. Your response should only include the answer. Do not provide any further explanation. 
\\ Sentence: \{\{sentence\}\}
\\ Fact:  \{\{incorrect-fact\}\}
\\ Answer:
\\\hline
\textbf{Step 2. Few-shot prompt} \\\hline
You are a highly intelligent assistant. You will carefully follow the instructions. Rephrase the following fact to be ambiguous, allowing for various interpretations or obscuring specific details. Your response should only include the answer. Do not provide any further explanation. Here are some examples, complete the last one: 
\\\\Fact: A third of Gen Zs believe the economic situation in their countries will improve over the next year. 
\\Answer: Many believe the economic situation in their countries will improve in the near future. 
\\\\Fact: Archaeologists have found evidence suggesting that the ancient ruins reveal details about early civilizations. 
\\Answer: It is believed that the ancient ruins reveal details about early civilizations. 
\\\\Fact: Senator Smith discussed foreign policy during his election campaign, and subsequently during his victory speech at the State Convention Center 
\\Answer: Senator Smith often discussed foreign policy in his speeches. \\\\Fact:  \{\{incorrect-fact\}\}
\\Answer: 
\\\hline
\textbf{Step 2. Few-shot prompt with context} \\\hline
You are a highly intelligent assistant. You will carefully follow the instructions. The user will provide a sentence containing a fact. Rephrase the given fact to be vague or ambiguous, allowing for various interpretations or obscuring specific details. Your answer should be still relevant to the sentence. Your response should only include the answer. Do not provide any further explanation. Here are some examples, complete the last one:
\\\\Sentence: A third of Gen Zs and millennials believe the economic situation in their countries will improve over the next year.
\\Fact: A third of Gen Zs believe the economic situation in their countries will improve over the next year.
\\Answer: Many believe the economic situation in their countries will improve in the near future.
\\\\Sentence: Archaeologists have found evidence suggesting that the ancient ruins reveal details about early civilizations.
\\Fact: Archaeologists have found evidence.
\\Answer: The evidence has been found.
\\\\Sentence: Senator Smith discussed foreign policy during his election campaign, and subsequently during his victory speech at the State Convention Center
\\Fact: Senator Smith discussed foreign policy during his victory speech.
\\Answer: Senator Smith often discussed foreign policy in his speech.
\\\\Sentence: \{\{sentence\}\}
\\Fact:  \{\{incorrect-fact\}\}
\\Answer: \\\hline
\end{tabular}
}
\caption{Four prompts tested for step 2.}~\label{tab:4prompts_step2}
\end{table*}

\begin{table*}[h]
\small
\begin{tabular}{p{\linewidth}}
\hline
\textbf{Step 3 - Aggregation prompt} \\\hline
You are a highly intelligent assistant. You will carefully follow the instructions. \\The user will provide multiple facts separated by a semicolon. Combine these facts into a single sentence. Ensure the sentence matches the structure of the reference. Your response should only include the answer. Do not provide any further explanation.
\\Fact:  \{\{correct fact\}\}; \{\{ambiguous fact\}\}; ...
\\Reference: \{\{original-sentence\}\}
\\Answer:
\\\hline
\end{tabular}
\caption{The aggregation prompt for step 3.}~\label{tab:aggregation}
\end{table*}

\begin{table*}[h]
\small
\begin{tabular}{p{\linewidth}}
\hline
\textbf{Steps 2 and 3 Combined - Iterative prompt} \\\hline
You are a highly intelligent assistant. You will carefully follow the instructions. The user will provide a sentence containing a fact that \{\{incorrect-fact\}\}. Your task is to rephrase the sentence to make the fact ambiguous, allowing for multiple interpretations or hiding specific details. Ensure that the revised sentence still conveys the other facts in the original sentence. If the sentence does not contain the fact, do not change the sentence. Your response should only include the answer. Do not provide any further explanation. 
\\Sentence:  \{\{sentence\}\}
\\\hline
\end{tabular}
\caption{A prompt that combines both step 2 and step 3 in the ambiguity approach. This prompt was called iteratively, with the output sentence that has been produced in the previous iteration as an input, along with a new incorrect fact.}~\label{tab:prompt_combined}
\end{table*}

\begin{table*}[h]
\small
\begin{tabular}{p{\linewidth}}
\hline
\textbf{Coherence}: Do all the sentences in the response collectively fit together and sound natural? (Context: Question) \\\hline
\begin{enumerate}
    \item The response is not coherent at all.
    \item The response is slightly coherent.
    \item The response is somewhat coherent.
    \item The response is mostly coherent.
    \item The response is very coherent.
\end{enumerate}
\\\hline
\textbf{Relevance}: Is the response relevant to the question? (Context: Question) \\\hline
\begin{enumerate}
    \item The response is not relevant to the question at all.
    \item The response is slightly relevant to the question.
    \item The response is somewhat relevant to the question.
    \item The response is mostly relevant to the question.
    \item The response is very relevant to the question.
\end{enumerate}
\\\hline
\textbf{Conciseness}: Is the response concise without extraneous information? (Context: Question) \\\hline
\begin{enumerate}
    \item The response is not concise at all.
    \item The response is slightly concise.
    \item The response is somewhat concise.
    \item The response is mostly concise.
    \item The response is very concise.
\end{enumerate}
\\\hline
\textbf{Completeness}: Is the response a complete answer to the question? (Context: Question) \\\hline
\begin{enumerate}
    \item The response is not complete at all.
    \item The response is slightly complete.
    \item The response is somewhat complete.
    \item The response is mostly complete.
    \item The response is very complete.
\end{enumerate}
\\\hline
\textbf{Correctness}: Is the response consistent with the reference paragraph without hallucination? (Context: Question, Wikipedia) \\\hline
\begin{enumerate}
    \item The response is not consistent with the reference at all.
    \item The response is slightly consistent with the reference.
    \item The response is somewhat consistent with the reference.
    \item The response is mostly consistent with the reference.
    \item The response is very consistent with the reference.
\end{enumerate}
\\\hline
\end{tabular}
\caption{Evaluation rubric for LLM-as-a-judge}~\label{tab:evaluation_prompt}
\end{table*}

\end{document}